# A Novel Modulation Scheme Based on the Kramers–Kronig Relations for Optical IM/DD Systems

Xiaohe Dong, Kuokuo Zhang, Jiarui Zhang, Baoyin Yang, Caiming Sun, *Senior Member, IEEE*

*Abstract*—The ever-growing demand for higher data rates in optical communication systems necessitates the development of advanced modulation formats capable of significantly enhancing system performance. In this work, we propose a novel modulation format derived from the Kramers–Kronig relations. This scheme effectively reduces the complexity of digital filtering and alleviates the demands on the digital-to-analog converter, offering a practical solution for high-speed optical communication. The proposed modulation format was rigorously validated through experimental investigations using an optical wireless link. The results demonstrate a notable improvement in bit error rate (BER) performance and receiver sensitivity compared to PAM-4 and CAP-16 modulation schemes, with enhancements of 0.6 dB and 1.5 dB in receiver sensitivity, respectively. These improvements enable higher data transmission rates, positioning the Kramers–Kronig relations-based modulation format as a promising alternative to existing modulation techniques. Its potential to enhance the efficiency and capacity of optical communication systems is clearly evident. Future work will focus on extending its application to more complex scenarios, such as high-speed underwater optical communication systems, where advanced modulation formats are critical for overcoming bandwidth limitations.

## I. INTRODUCTION

OPTICAL communication systems are at the heart of modern telecommunications, enabling the high-speed data transmission necessary to support the rapidly growing global network infrastructure. As the demand for data rates continues to escalate, particularly within data centers and backbone networks, the choice of modulation formats plays a critical role in determining the efficiency and capacity of optical links. Intensity-modulated direct detection (IM-DD) systems, with their low-cost configurations, are well-suited for Ethernet, especially as speeds and the number of lanes increase through wavelength division multiplexing (WDM) and parallel single-mode fiber (PSM).

Studies have shown that increasing the number of lanes can achieve higher overall transmission capacity, even at a relatively lower per-lane speed[1-6]. While IM-DD systems offer economic advantages, the emerging Ethernet standards must aim to reduce the number of multiplexing channels, thereby enabling fewer lanes with higher per-lane transmission rates. Therefore, advanced modulation formats have garnered substantial attention from both the research community and industry as promising approaches of reducing the number of required lanes while simultaneously increasing overall link capacity. M-PAM [7], DMT [8,9] and CAP [10-12] has been reported in high speed optical IM/DD communication system.

Turning to optical wireless communication (OWC) systems, bandwidth limitations arising from both the transmitter and receiver, coupled with inter-symbol interference (ISI), continue to represent significant constraints on overall link capacity. Advanced modulation schemes have also emerged as promising approaches to address these limitations and improve link capacity. Several studies have demonstrated the potential of these advanced techniques. For instance, in [13], PAM-4 modulation combining with a decision feedback equalizer was utilized on VCSEL-based FSO link cutting off the bandwidth to half compared to that of NRZ modulation. Similarly, CAP modulation is used to achieve 4.5 Gb/s [14] and a pre-equalized PAM-4 scheme enabled 1 Gb/s for free-space visible light communication [15]. Moreover, orthogonal frequency-division multiplexing (OFDM) was leveraged to mitigate bandwidth limitations, achieving a transmission rate of 2.32 Gb/s aided by two-stage linear software equalizer [16]. Additional research focusing on the application of advanced modulation schemes to enhance OWC link performance can be found in [17-22]. These developments underscore the substantial potential of combining advanced modulation formats with equalization techniques to significantly improve the capacity of OWC links.

Despite these advancements, traditional advanced modulation formats still encounter several limitations. PAM, while simple and cost-effective, is prone to baseband wandering [23], which can negatively impact performance at

Manuscript received XX September 2024; revised XX XXX 2024; accepted XX XXX 2024; This work was supported in part by the National Natural Science Foundation of China under Grant 62175120, in part by the Shenzhen Science and Technology Program under Grant JCYJ20220818103011023, and in part by Longgang District Shenzhen's "Ten Action Plan" for Supporting Innovation Projects under Grant LGKCSDPT2024002. (Corresponding author: Caiming Sun.)

Xiaohe Dong, Kuokuo Zhang, Jiarui Zhang, Baoying Yang, and Caiming Sun are with the Shenzhen Institute of Artificial Intelligence and Robotics for Society (AIRS), School of Science and Engineering, the Chinese University of Hong Kong, Shenzhen 518172, China (email: dongxiaohe@cuhk.edu.cn; zhangkuokuo@cuhk.edu.cn; z2323036076@163.com; 879339863@qq.com; cmsun@cuhk.edu.cn).

higher data rates. OFDM/DMT, although resilient to frequency-selective fading, necessitates the application of Hermitian symmetry to convert the signal into the real value [8], which in turn reduces spectral efficiency and also suffers from the high peak-to-average power ratio (PAPR) [24]. CAP modulation, while advantageous in minimizing baseline wander and maintaining spectral efficiency, complicates system design due to its reliance on multiple digital filters [25].

To address these challenges, we propose a novel modulation format based on the Kramers–Kronig (KK) relations. This new scheme is specifically designed to mitigate the shortcomings of existing formats, offering several key benefits. In comparison to OFDM, the proposed format enhances spectral efficiency by obviating the need for Hermitian symmetry. As a single-carrier based modulation scheme, it does not generate high PAPR. Unlike CAP, it only requires one digital filter and therefore simplifies system implementation by reducing the number of required digital filters, thereby decreasing computational complexity. Additionally, the proposed scheme imposes lower demands on the Digital-to-Analog Converter (DAC) compared to PAM, making it a more efficient solution for high-speed optical communication systems.

This paper presents the first thorough evaluation of this Kramers–Kronig (KK) relation-based modulation format for optical links. The remainder of this paper is organized as follows: Section II provides an in-depth exploration of the proposed modulation scheme and its theoretical foundation. Section III presents the simulation results over a bandwidth limited link. Section IV details the experimental setup used to assess the performance of the new format. Section V discusses the experimental results, highlighting significant improvements in link performance, ease of implementation, and reduced hardware requirements. Finally, Section VI concludes the paper and outlines potential directions for future research.

## II. THEORETICAL ANALYSIS

The Kramers–Kronig (KK) relations are pervasive in various domains of physics and engineering, with applications spanning multiple areas [26–30]. In the context of direct-detection coherent receivers, their relevance arises from a fundamental property of minimum-phase signals, as detailed in [32,33]. Specifically, in [31], a Kramers–Kronig-based coherent receiver is proposed to recover complex data information directly from the current of the optical detector. In this paper, we take a different approach by constructing a minimum-phase signal in an intensity-based system and extracting its phase information from the signal itself. This method allows simultaneous transmission of both in-phase and quadrature-phase components. The central concept of our scheme is the identification of a minimum-phase condition, which ensures that the phase of the signal can be uniquely recovered from its intensity through the Hilbert transform. We define the baseband signal s(t) as a complex data-carrying signal, whose spectrum is constrained within the frequency range between −B/2 and B/2, and consider a single-sideband signal of the form:

$$h(t) = A + s(t)\exp(i\omega t) \quad (1)$$

where A is constant. It has been established that h(t) is a minimum-phase signal if and only if the winding number of its time trajectory in the complex plane is zero, meaning that it does not encircle the origin [32]. This condition is satisfied when A is sufficiently large. Under the minimum-phase condition, the phase $\phi(t)$ of the transmitted signal is uniquely related to the Hilbert transform, as described by the following expression:

$$\phi(t) = \frac{1}{\pi} p.v. \int_{-\infty}^{\infty} dt' \frac{\log|h(t')|}{t - t'} \quad (2)$$

where p.v. stands for principal value, |h(t)| is the modulus of the signal h(t). The equation (2) is one of the two KK relations existing between $\phi(t)$ and $\log|h(t)|$. Therefore, the minimum phase signal can be expressed by:

$$h(t) = |h(t)|\exp(\phi(t)) \quad (3)$$

So the phase of the signal h(t) can be uniquely determined by its modulus only. The |h(t)| can be expressed:

$$|h(t)| = |h(t)\exp(-i\omega t)| \quad (4)$$

$$= |A\exp(-i\omega t) + s(t)|$$

The baseband signal s(t) includes the information transmitted and therefore let s(t) equal:

$$s(t) = I(t) - Q(t) * i \quad (5)$$

where $I(t)$ are the in-phase transmitted symbols and $Q(t)$ are the quadrature-phase channel transmitted symbols. Therefore, the modulus of the minimum phase signal h(t) can be expressed:

$$|h(t)| = |A\cos(\omega t) - A\sin(\omega t) * i + I(t) - Q(t) * i| \quad (6)$$

Combining the real and imaginary part of the equation, the transmitted signal $|h(t)|$ can be further expressed:

$$|h(t)| = \sqrt{[I(t) + A\cos(\omega t)]^2 + [Q(t) + A\sin(\omega t)]^2} \quad (7)$$

According to equation (1), the baseband signal s(t) can be demodulated by:

$$s(t) = (|h(t)|\exp(\phi(t)) - A) * exp(-i\omega t) \quad (8)$$

Therefore, its in-phase and quadrature phase symbols can be expressed as:

$$I(t) = \Re[(|h(t)|\exp(\phi(t)) - A)\exp(-i\omega t)] \quad (9)$$

$$Q(t) = -\Im[((|h(t)|\exp(\phi(t)) - A))\exp(-i\omega t)] \quad (10)$$

where $\Re[.]$ and $\Im[.]$ represents the real part and imaginary part of the expression. Therefore, the transmitted data bits are first parallelized into two streams and can be encoded using a multilevel encoder. Subsequently, a Nyquist pulse shaping filter is applied to enhance spectral efficiency. The Nyquist filter is characterized by a rolloff factor α=0.1, a filter length of Nsym=30, and a frequency of $\frac{R_S(1+\alpha)}{2}$, where $R_s$ is the symbol rate. It enables the construction of the intensity-based signal y(t):

$$y(t) = \sqrt{[I(t) + A\cos(\omega t)]^2 + [Q(t) + A\sin(\omega t)]^2} \quad (11)$$

where $\omega$ denotes the frequency corresponding to $\pi R_S(1+\alpha)$. The amplitude $A$ is critical for ensuring that the signal meets the minimum phase condition. The phase of the signal can then be demodulated using the Hilbert transform. Fig. 1 illustrates the modulation and demodulation process of this novel scheme, which is based on the KK relation. To satisfy the minimum phase condition, it is essential to select an optimal value for A to ensure accurate signal demodulation. Fig. 2 presents theoretical demodulation diagrams of KK-16 for various values of A. The analysis shows that for A=0, A=3, and A=5, the reconstructed signal fails to meet the minimum phase condition, leading to incomplete or erroneous constellation recovery. In contrast, when A=8, A=10, and A=20, the minimum phase condition is satisfied, resulting in successful constellation recovery. These findings highlight the significance of choosing appropriate parameter values. Furthermore, Fig. 3 depicts the error vector magnitude (EVM) as a function of A, providing a quantitative assessment of the demodulation performance across different values of A. The EVM is defined:

$$\text{EVM}(\%) = \frac{\sqrt{\frac{1}{N}\sum_{k=1}^{N}|S'(k) - S(k)|^2}}{\sqrt{\frac{1}{N}\sum_{k=1}^{N}|S(k)|^2}} * 100\% \quad (12)$$

The error vector magnitude (EVM), which quantifies the deviation between the ideal demodulation symbols S(k) and the received symbols S′(k), is observed to decrease as the parameter A increases. Notably, the EVM reaches a saturation point when A exceeds 20, highlighting the significant impact of A on demodulation accuracy. Additionally, A influences the average value of the modulated signal, as shown in Fig. 4(a), where the average value increases with larger values of A. Importantly, the bias point of the modulated signal aligns with its average value, a relationship further illustrated in Fig. 5. The actual modulated signal that drive the MZM is:

$$y'(t) = y(t) - \overline{y(t)} \quad (13)$$

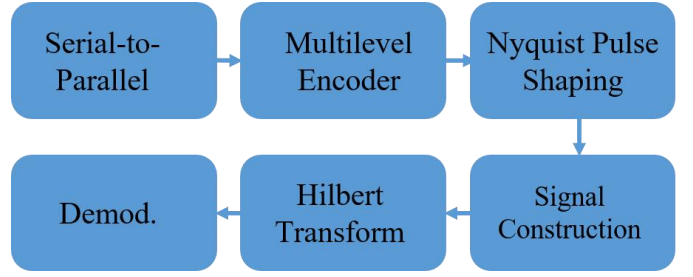

**Fig. 1.** Schematic representation of the modulation and demodulation principles underlying the proposed Kramers-Kronig based scheme (Demod.: Demodulation).

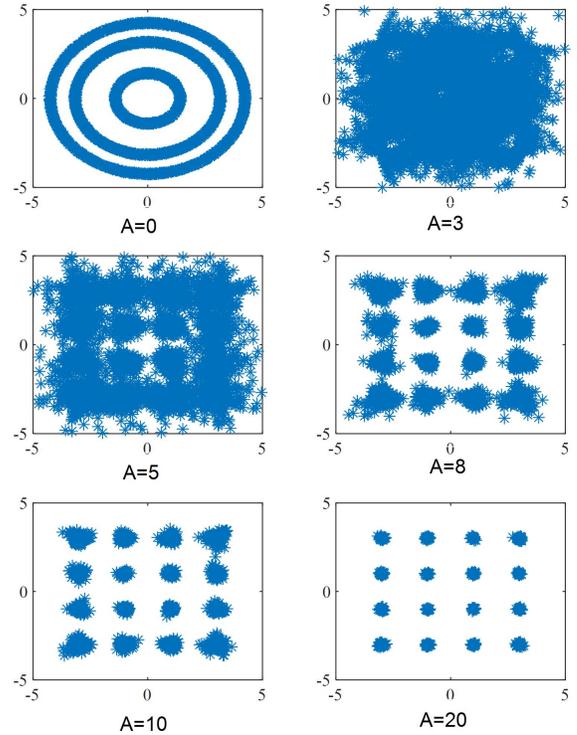

**Fig. 2.** Theoretical demodulation diagrams illustrating the effect of varying the parameter $A$ on constellation recovery.

where $\overline{y(t)}$ represents the average value of the y(t). Another critical characteristic of the modulated signal is its peak-to-peak value, as depicted in Fig. 4(b). For A values less than 10, the peak-to-peak value increases with rising A. However, when A reaches or exceeds 10, the peak-to-peak value stabilizes at approximately 13.5, indicating that further increases in A have no significant effect on this parameter. This behavior carries important implications for the performance of the optical communication link. At higher values of A, assuming identical optical link and components parameters, the Euclidean distance between symbols remains constant. This ensures that the optical system maintains its power efficiency. Consequently, increasing A results in improved demodulation performance without compromising power efficiency, which is a key advantage for optimizing system performance.

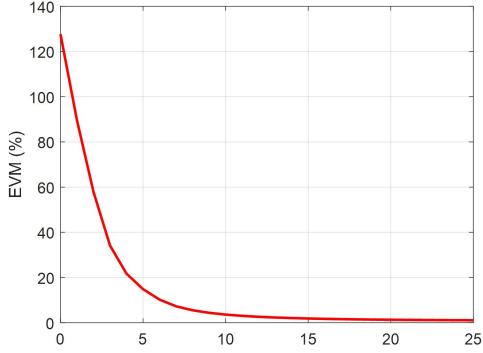

**Fig. 3.** Error vector magnitude versus value of A for the KK based modulation scheme

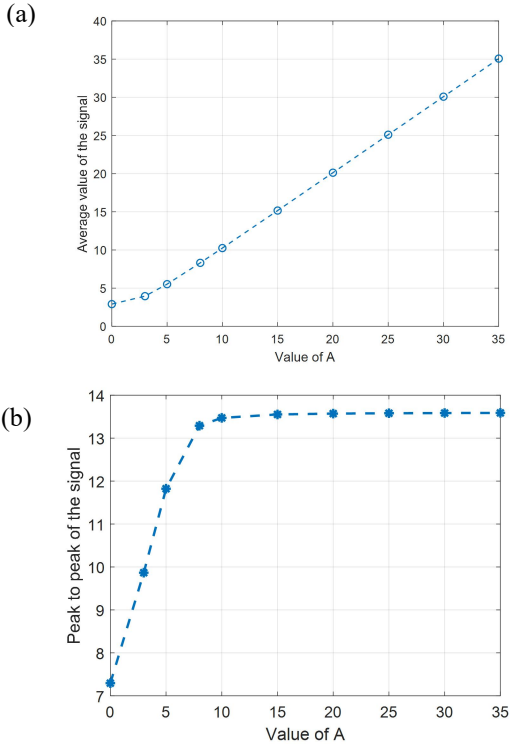

**Fig. 4.** (a) Average value and (b) peak-to-peak value of the signal as a function of A for the KK based modulation scheme.

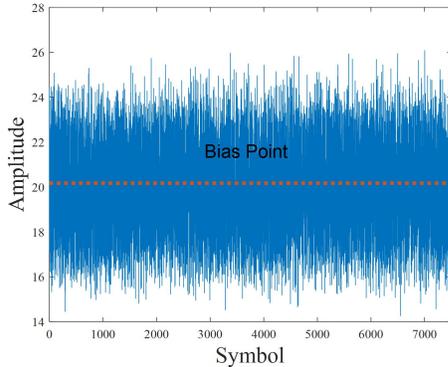

**Fig. 5**. Bias point of the modulated signal for A=20

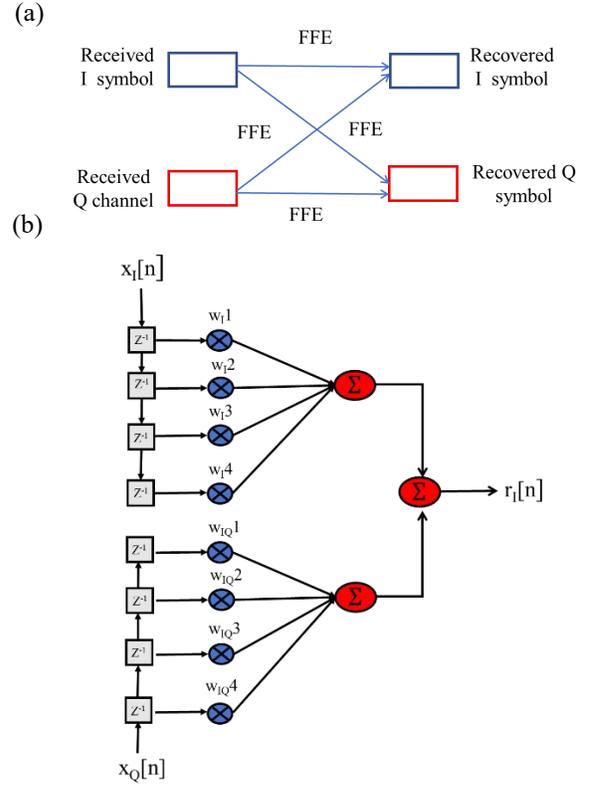

**Fig. 6.** The structure of the employed equalizer: (a) overall structure, (b) detailed structure for the I- channel

III. SIMULATION RESULTS OVER A BANDWIDTH LIMITED LINK

In this section, we present a theoretical analysis of the KK-based modulation scheme and investigate its performance over a bandwidth-limited optical link. To model the overall link response, we employ a Gaussian low-pass filter, which represents the limited bandwidth of the link. To address the inter-symbol interference (ISI) issue that arise due to the bandwidth limitation, we incorporate an equalizer into the system before demodulation. For the purpose of analysis, we set the 3 dB bandwidth of the optical link to 1 GHz. Fig. 6 depicts the structure of the equalizer used to recover the transmitted symbols from both the I- and Q-channels, which has been proposed in [10-12] Specifically, Fig. 6(a) presents the overall structure of the equalizer, while Fig. 6(b) provides a detailed illustration of the equalizer structure for the I-channel. It is important to note that the Q-channel has the same detailed structure as the I-channel.

The mathematical expression of the equalizer is shown:

$$r_I(n) = \sum_{m=0}^{M} x_I(n-m)w_I(m) + \sum_{l=0}^{L} x_Q(n-l)w_{IQ}(m)$$

$$r_Q(n) = \sum_{m=0}^{M} x_Q(n-m)w_Q(m) + \sum_{l=0}^{L} x_I(n-l)w_{QI}(m)$$

In this setup, the equalizer consists of a total of 16 taps, comprising 10 taps (M=10) from the main channel and 6 taps

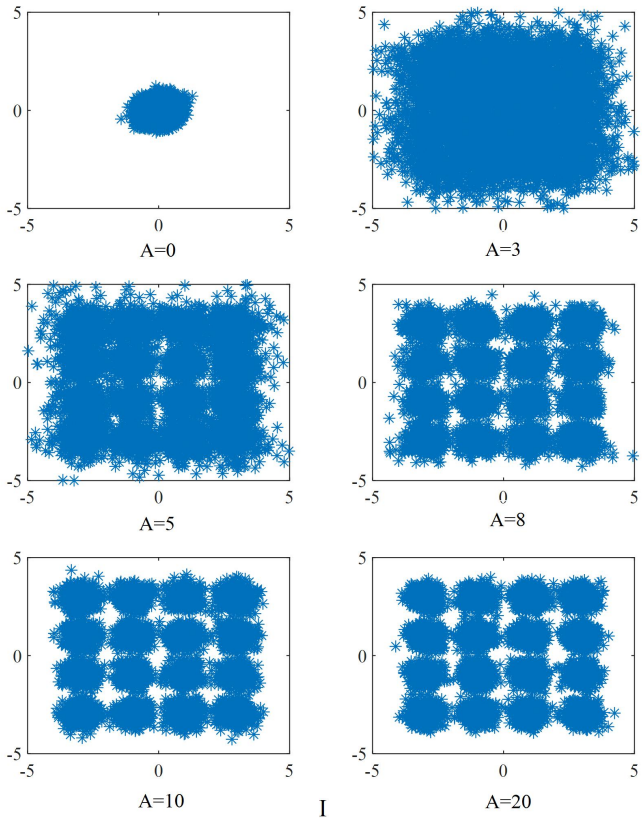

**Fig. 7.** Demodulation results after equalization for 12 dB OSNR at 1Gbaud transmission rate

(L=6) from the cross-talk channel. Figure 7 presents the demodulation results with varying values of A after equalization, at an optical signal-to-noise ratio (OSNR) of 12 dB and a transmission baud rate of 1 Gbaud. The constellation diagrams progressively become clearer as the value of A increases.

Fig. 8 illustrates the bit error rate (BER) performance as a function of the parameter A when the symbol rate is set to 1 GBaud. when A=0, the symbols cannot be demodulated correctly, even with an increasing optical signal-to-noise ratio (OSNR). Furthermore, the BER performance tends to saturate once A exceeds a value of 20. The plot clearly shows that the BER curves for A=20 and A=25 overlap, indicating that the link performance reaches its saturation point when A exceeds 20.

Fig. 9 presents the BER performance with A=20 for different symbol rates after equalization. As the symbol rate increases, the system performance degrades due to the limited bandwidth, which introduces greater inter-symbol interference (ISI) and causes a deterioration of the received signal. Fig. 10 illustrates the demodulation results for different baud transmission rate with the same equalizer length (10 taps from the main channel and 6 taps from the other channel) at 12 dB OSNR. It is clear to show that the link performance deteriorates with the increasing data transmission rate as the larger ISI effect imposed on the link with the higher data rate.

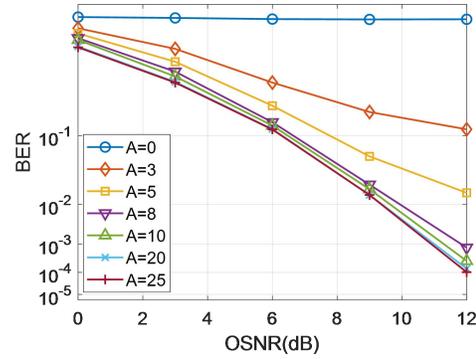

**Fig. 8.** BER value against different value of OSNR for 1Gbaud transmission rate

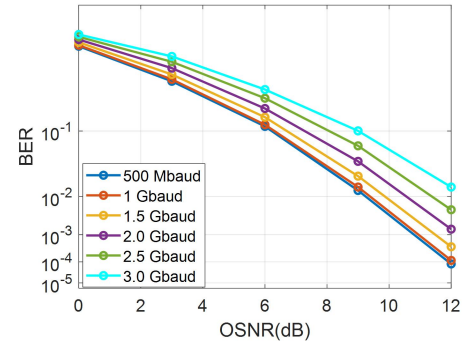

**Fig. 9.** BER performance for A=20 at different data transmission rate

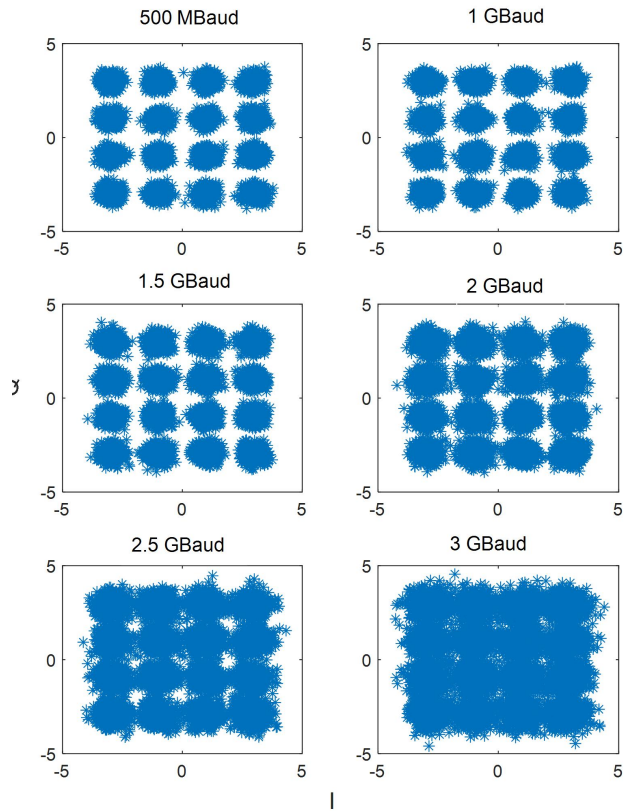

**Fig. 10.** Demodulation results under different data transmission rate at OSNR=12 dB for A=20

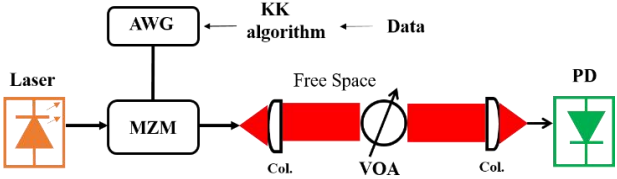

**Fig. 11.** Experimental setup used in the data transmission test.

## IV. EXPERIMENTAL CONFIGURATION

In our experimental framework, we evaluate the performance of the proposed modulation scheme. Data transmission measurements were conducted over a 50 cm optical wireless communication (OWC) link, as depicted in Fig. 11. At the transmitter end, a 2.5 GSa/s arbitrary waveform generator (AWG, SIGLENGT SDG7102A) with a 1 GHz analog bandwidth was employed to generate the high-speed modulated signal. The modulation was implemented using one arm (MZM A) of an I-Q modulator (SUMITOMO T.SBXH1.5-20PD-ADC) with a 20 GHz analog bandwidth. The MZM A was biased at 0.5 v, with the peak-to-peak voltage of the modulated signal set at 1 volt to ensure linear operation, while the other arm (MZM B) was not utilized.

Each data pattern consisted of 30,000 randomly generated bits, and 10 such patterns were used for bit error rate (BER) testing. The data was first encoded into the desired symbol format, and this encoded data stream was used to drive one channel output of the AWG, which was connected to the corresponding arms of the I-Q modulator.

The modulated optical signal was then coupled into free space through an optical collimator. The transmitted beam was then collected by an objective lens and coupled on the 0.2 mm-diameter active area of a free-space photodetector. A free-space variable optical attenuator (VOA) was employed to precisely control the received optical power (ROP) of the photodetector. This photodetector, with a 3 dB bandwidth of 1.5 GHz, converted the received optical signal into its electrical equivalent. The captured waveform was then recorded by a 50 GHz sampling oscilloscope (Tektronix MS064B) with a 2.5 GHz analog bandwidth. Subsequent offline signal processing, including demodulation, equalization, and BER computation, was performed using MATLAB.

## V. EXPERIMENTAL RESULTS

In this section, we present a comprehensive analysis of the experimental results obtained from the implementation of the proposed modulation scheme. The experiments were conducted over an optical wireless communication (OWC) link to assess the performance of the novel modulation format, with a focus on comparing it to the PAM-4 and CAP-16 modulation. Due to the limitations of the arbitrary waveform generator (AWG) used in this setup (2.5 GSa/s), the objective is not to push the boundaries of data rate, but rather to experimentally validate the underlying principles of the proposed modulation scheme.

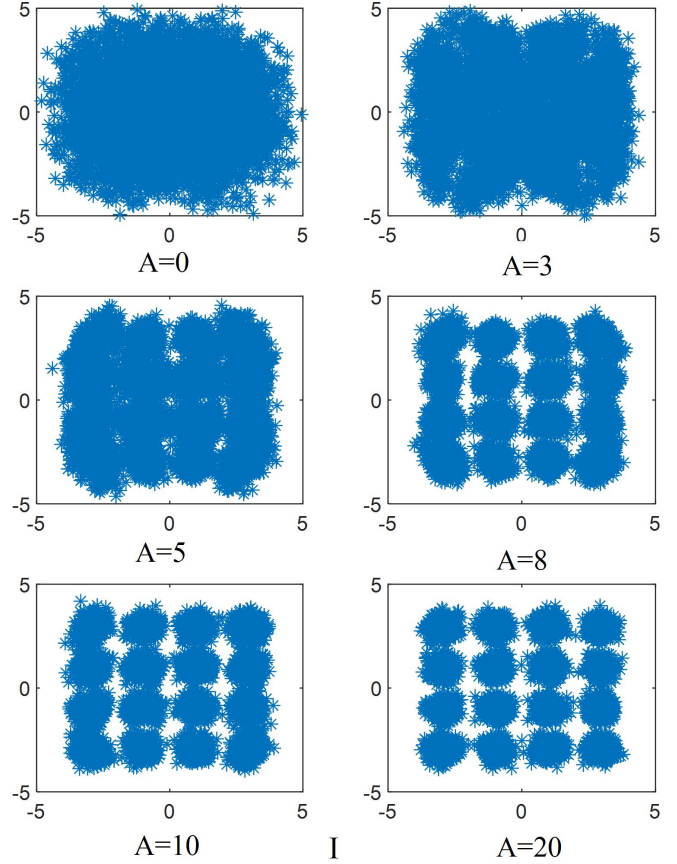

**Fig. 12.** Constellation diagrams of the proposed KK based modulation scheme at 2.5 Gb/s with a ROP of 1 μW, demonstrating the effect of varying parameter A on signal clarity.

### A. Impact of Parameter A on System Performance

Fig. 12 illustrates the constellation diagrams of the proposed modulation scheme at a data rate of 2.5 Gb/s, under ROP level of 1 μW. The constellations were captured following the application of a novel equalizer structure, as discussed in Section III. The results reveal a progressive improvement in constellation clarity as the parameter $A$ increases, as predicted by the theoretical illustration in Fig. 2. Specifically, the constellations corresponding to A=8, A=10, and A=20 demonstrate well-defined symbol separation, indicating successful demodulation and robust performance under the tested conditions.

The BER performance of the OWC link, as a function of the parameter A, is presented in Fig. 13. The experimental data indicate that the proposed modulation scheme achieves a BER below the hard-decision forward error correction (HD-FEC) threshold of $3.8 \times 10^{-3}$ for $A=8$ and higher. In contrast, the configurations with $A=3$ and $A=5$ fail to meet the HD-FEC threshold, further corroborating the necessity of optimal parameter selection to ensure reliable transmission. It is noteworthy that the BER plot for A=0 was omitted, as the corresponding constellation diagrams clearly indicate a failure in signal demodulation. Furthermore, the receiver sensitivity

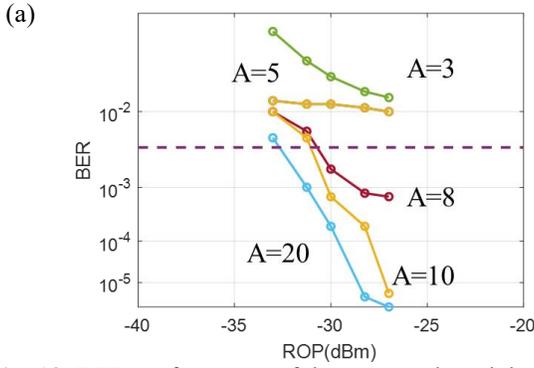

**Fig. 13.** BER performance of the proposed modulation scheme as a function of the parameter *A* at 2.5 Gb/s data transmission.

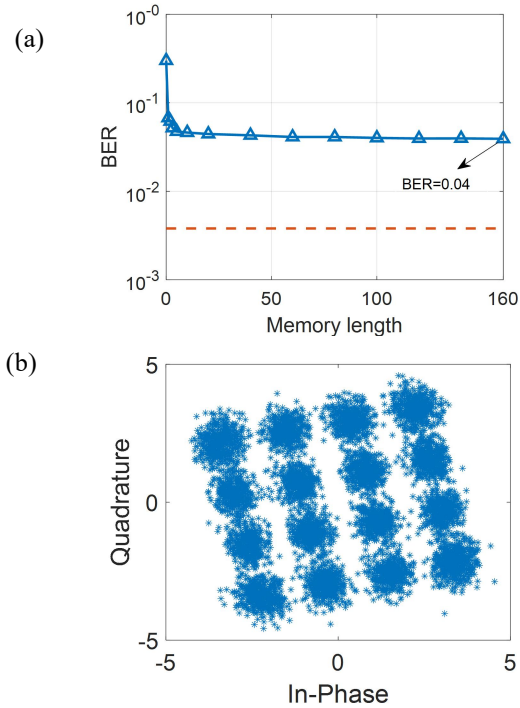

**Fig. 14.** (a) BER performance with increased FFE tap length on the main channel (b) demodulation results with 100 FFE taps on the main channel for 2.5 Gb/s data transmission at ROP of 1μW

required to achieve the HD-FEC limit was observed to be approximately -31.5 dBm, -32 dBm, and -33 dBm for *A=8*, *A=10* and *A=20*, respectively. Fig. 14(a) demonstrates the effect of equalizer memory length on bit error rate (BER) performance for I- channel when A=20 and the received optical power (ROP) is 1 μW for a data transmission rate of 2.5 Gb/s. The performance reaches saturation at approximately 25 taps, with further increases in the number of taps yielding only marginal improvements. Fig 14(b) presents the demodulation results, clearly illustrating that the constellation diagram experiences rotation due to the crosstalk effect from the adjacent channel. This phase rotation can be effectively compensated for by applying a small number of taps on the crosstalk channel.

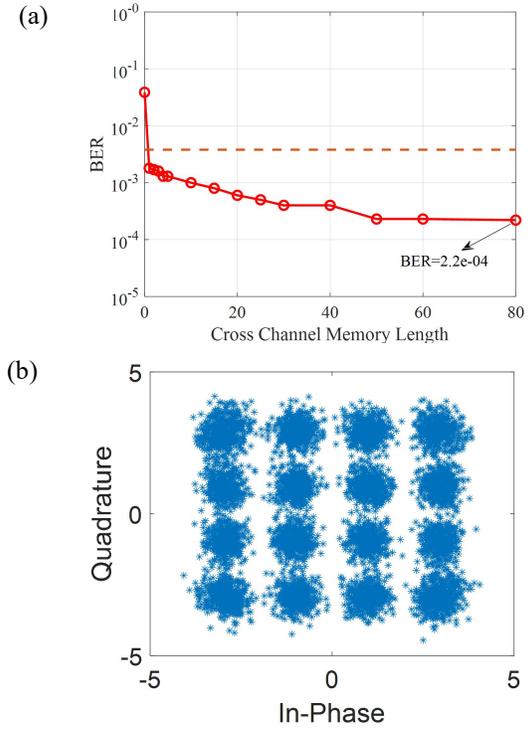

**Fig. 15.** (a) Bit error rate (BER) performance with increased FFE tap length on the crosstalk channel, while the main channel tap length is fixed at 100 taps. (b) Demodulation results with 60 FFE taps on the crosstalk channel for 2.5 Gb/s data transmission at ROP of 1 μW.

Fig. 15(a) shows the impact of the FFE memory length on the crosstalk channel with respect to the BER performance of the link. The corresponding demodulation map with 60 taps applied to the crosstalk channel is also presented in Fig. 15(b). Even with just a single tap, the phase rotation effect is effectively mitigated, demonstrating the efficiency of the equalizer in counteracting the crosstalk-induced distortion.

### B. Comparative Analysis with PAM-4 modulation and CAP-16 modulation

To further validate the efficacy of the proposed modulation scheme, we compared its BER performance against the PAM-4 and CAP-16 modulation at 2.5 Gb/s. Fig. 16 shows the demodulation results of CAP-16 and the eye diagram of the PAM-4 signal at ROP of 1 μW. Furthermore, the comparison, illustrated in Fig. 17, reveals that the proposed scheme outperforms PAM-4 and CAP-16 by a margin of 0.6 dB and 1.5 dB in receiver sensitivity. Besides, the proposed modulation scheme provides a simpler implementation with one digital filter to operate the Hilbert transform than the CAP modulation which requires 4 digital filters. In addition to this improvement, the proposed modulation format also lowers the required sampling rate of the Digital-to-Analog Converter (DAC), thereby enabling more efficient and practical implementations in high-speed optical communication systems, which will be discussed in the next section. This dual advantage—enhanced receiver sensitivity and reduced DAC requirements—underscores the superiority of the proposed modulation format in achieving higher data rates. Figure 18 presents the BER performance of three modulation schemes—PAM-4, the KK scheme, and CAP-16—at various data rates, with a received optical power (ROP) of 500 nW. The results indicate that the KK scheme consistently outperforms both the PAM-4 and CAP-16 schemes across all data rates. Figure 19 displays the demodulation constellation diagrams for the KK scheme at an ROP of 500 nW, illustrating its stable performance even under low ROP conditions.

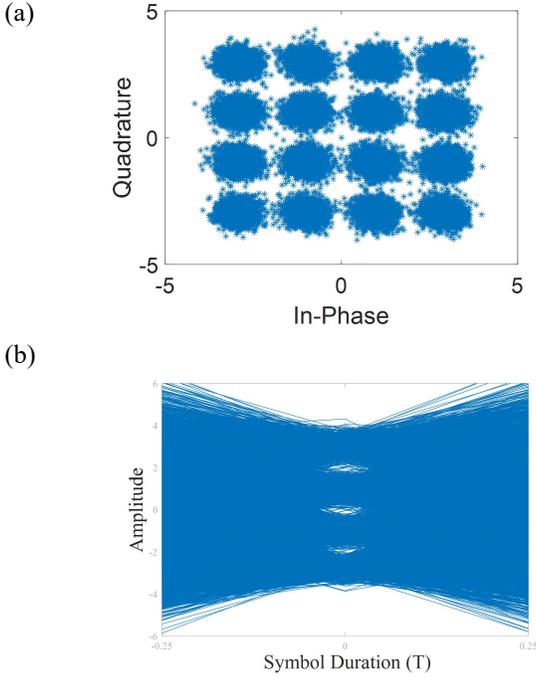

**Fig.16.** Demodulation results for the CAP-16 modulation scheme and eye diagram of the PAM4 modulation at 2.5 Gb/s under ROP of 1 μW

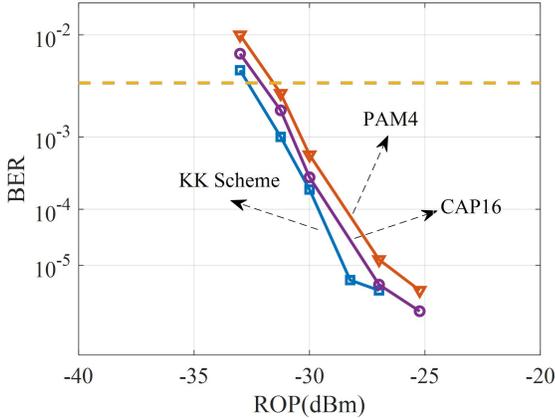

**Fig. 17.** Comparative analysis of BER performance among the proposed KK-based modulation scheme, CAP-16 and PAM-4 modulation at 2.5 Gb/s, showcasing 0.6 dB and 1.5 dB improvement in receiver sensitivity over CAP-16 and PAM-4, respectively.

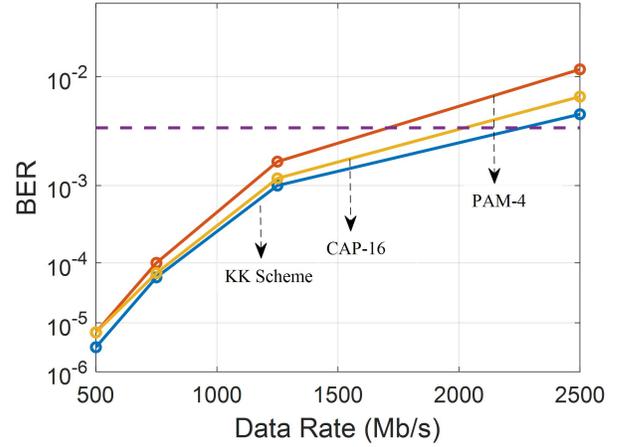

**Fig 18.** BER plot of KK scheme at various data rate under 500 nW ROP

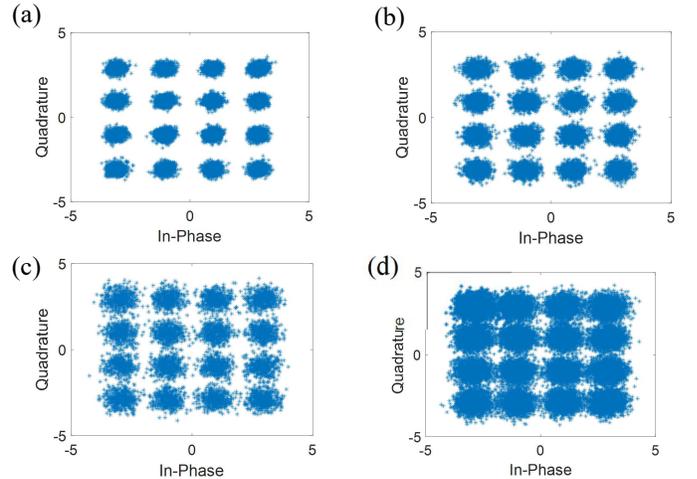

**Fig. 19.** Demodulation results for KK based modulation under data rates of: (a) 500 Mb/s, (b) 750 Mb/s, (c) 1.25 Gb/s, (d) 2.5 Gb/s at ROP of 500 nW

*C. High-Speed Performance Evaluation*

The data transmission rate in the experiment is constrained by the maximum sampling rate of the AWG, which is 2.5 GSa/s. For the PAM-4 modulation scheme, adhering to the Nyquist sampling criterion necessitates a maximum symbol rate of 1.25 GSa/s, thereby limiting the achievable data rate to 2.5 Gb/s. In contrast, the proposed modulation scheme, when utilizing a 2.5 GSa/s sampling rate, achieves a symbol rate of 1.25 GSa/s, enabling a data transmission rate of 5 Gb/s. For the CAP-16 modulation scheme, in order to achieve a 5 Gb/s transmission rate, an oversampling ratio of 2 is required, with the symbol rate remaining consistent with that of the proposed scheme. However, insufficient oversampling results in significant distortion in the quadrature channel, which adversely affects the signal recovery and degrades link performance. Fig. 20 shows the 5 Gb/s data transmission for the CAP-16 modulation at 1.5 μW and 3 μW. It is clear that

the Quadrature channel symbols remain distorted even after the equalization process. Fig. 21

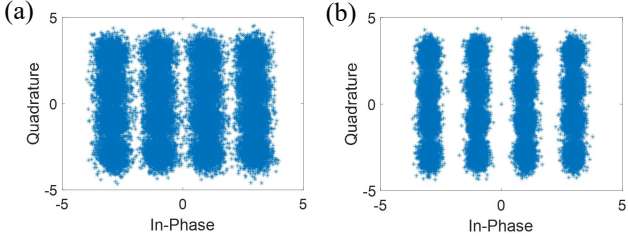

**Fig. 20.** Constellation diagrams at 5 Gb/s data transmission for CAP-16 (oversampling ratio=2) modulation under ROP level of: (a) 1.5 µW and (b) 3 µW.

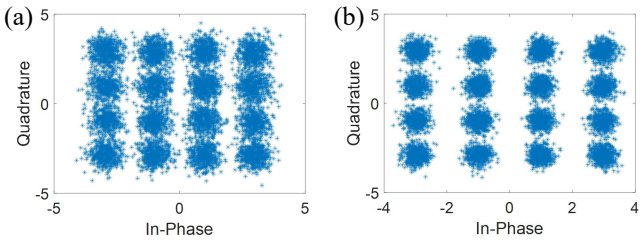

**Fig. 21.** Constellation diagrams at 5 Gb/s data transmission for KK scheme (A=20, oversampling ratio=2), under ROP levels of (a) 1.5 µW and (b) 3 µW, highlighting the robustness of the modulation scheme at higher data rates.

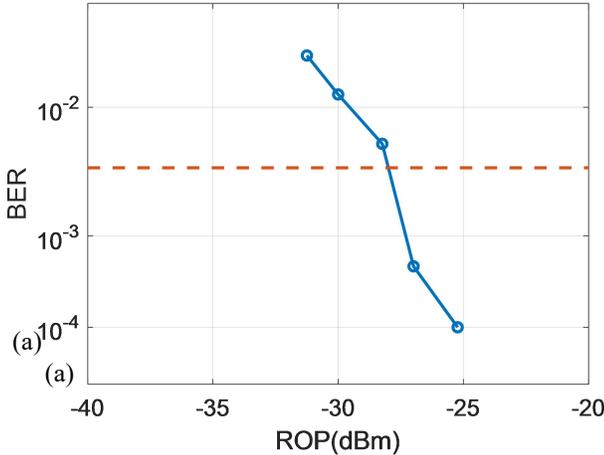

**Fig. 22**. BER performance for KK scheme at 5 Gb/s data transmission with the HD-FEC limit as a benchmark.

presents the constellation diagrams of the proposed modulation scheme at a data rate of 5 Gb/s, under ROP levels of 1.5 µW and 3 µW. The results exhibit a stable constellation pattern, even at the higher data rate, indicating the scheme's potential for scaling to more demanding transmission scenarios. Correspondingly, the BER performance at 5 Gb/s, shown in Fig. 22 demonstrates that the receiver sensitivity at the HD-FEC limit is approximately -27.5 dBm, further highlighting the modulation scheme's capacity for high-speed optical communication. These results also indicate the proposed scheme can ease the DAC sampling rate in comparison with the PAM and CAP modulation.

## VI. CONCLUSION

In this study, we have introduced a novel modulation format based on the KK scheme. Through an extensive theoretical framework and meticulous experimental validation, we have demonstrated that the proposed modulation format offers advantages in performance and system implementation simplicity, particularly within the context of intensity-modulated direct detection (IM-DD) systems. Specifically, in comparison with CAP modulation, it enables a simpler implementation than the CAP modulation by reducing the number of digital filters required.

Our experimental investigations, conducted using a OWC link, reveal the superior performance of the proposed modulation format. The experimental results indicate a 0.6 and 1.5 dB improvement in receiver sensitivity for 2.5 Gb/s data transmission test when compared to the PAM-4 and CAP-16 modulation scheme. The novel modulation format exhibits enhancement in receiver sensitivity, thereby enabling higher data transmission rates. Furthermore, the reduced demand on the DAC substantiates the potential of this modulation format to advance high-speed optical communication systems.

These findings substantiate the efficacy of the proposed modulation format as a compelling alternative to existing modulation techniques, with the potential to significantly augment the efficiency and capacity of optical communication links. Future research will focus on extending the applicability of this approach to more complex optical communication scenarios, including underwater optical communication systems. The robustness and adaptability of the proposed modulation format make it particularly promising for application in underwater optical wireless communication systems, where high-speed data transmission requires advanced modulation formats to overcome the challenges of limited bandwidth.

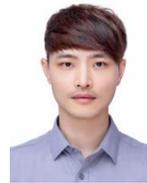

Xiaohe Dong, Ph.D., Assistant Researcher at the Shenzhen Institute of Artificial Intelligence and Robotics for Society (AIRS), received his B.S. ad M.S. degree in University of Liverpool and University of Cambridge in 2014 and 2016, respectively. He obtained his Ph.D. in Optical Communications from the University of Cambridge in 2021. His current research focuses on underwater optical communication, multimode fiber optic communication, wireless optical communication, and digital signal processing technologies, including channel equalization, artificial neural networks, and advanced modulation formats.

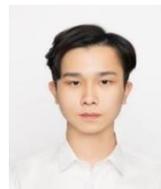

Kuokuo Zhang, Master's, Engineer at the Shenzhen Institute of Artificial Intelligence and Robotics for Society (AIRS). He graduated from Shenzhen University in 2022 with a Master's degree in Optical Engineering. His main research interests are in wireless underwater optical communication and all-optical communication devices.

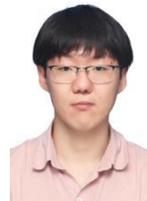

Jiarui Zhang, Master's, Engineer at the Shenzhen Institute of Artificial Intelligence and Robotics for Society (AIRS). He graduated with a Bachelor's degree in electronic and electrical engineering in University of Liverpool in 2021 and Master's degree in Optical Communications from the University of Bristol in 2024. His research direction includes photon device design, simulation, testing, algorithm optimization, and experimental verification.

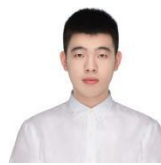

Baoyin Yang, Master's, currently pursuing a Master's degree at Guangzhou University, currently working as a visiting student at the Shenzhen

Institute of Artificial Intelligence and Robotics for Society (AIRS). His research direction is the application of grating couplers in optical phased arrays and their PT-symmetry optimization.

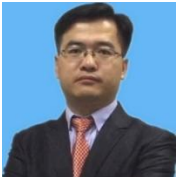

Caiming Sun (Senior Member, IEEE), Ph.D., received his B.S. and M.S. degrees from Beijing Normal University, Beijing, China, in 2002 and 2005, respectively, and his Ph.D. in Electronic Engineering from the Chinese University of Hong Kong (CUHK) in 2008. With over 10 years of R&D experience, supported by NSFC and HK ITF, his work covers optical communications, nanophotonics, nanofabrication, and wearable electronics. His current research interests include LiDAR technologies for robotics, optical wireless communications, and silicon photonics. Dr. Sun is currently a Research Associate Professor at the Shenzhen Institute of Artificial Intelligence and Robotics for Society (AIRS), School of Science and Engineering, The Chinese University of Hong Kong, Shenzhen, China.